\documentclass[12pt,preprint]{aastex}
\newcommand{\kms}{\mbox{km~s$^{-1}$}}

\newcommand{\aca}{$\alpha$~Cen~A}
\newcommand{\acb}{$\alpha$~Cen~B}
\newcommand{\eeri}{$\epsilon$~Eri}

\shorttitle{{\em FUSE} Survey of Coronal Lines}
\shortauthors{Redfield et al.}

\begin{document}

\title{A {\em Far Ultraviolet Spectroscopic Explorer} Survey of Coronal Forbidden Lines in Late-Type Stars}

\author{Seth Redfield\altaffilmark{1}, Thomas R Ayres\altaffilmark{2}, Jeffrey L. Linsky\altaffilmark{3}, Thomas B. Ake\altaffilmark{4}, A. K. Dupree\altaffilmark{5}, Richard D. Robinson\altaffilmark{4}, Peter R. Young\altaffilmark{5,6}}

\altaffiltext{1}{JILA, University of Colorado, Boulder, CO 80309-0440; sredfiel@casa.colorado.edu}
\altaffiltext{2}{CASA, University of Colorado, Boulder, CO 80309}
\altaffiltext{3}{JILA, University of Colorado and NIST, Boulder, CO 80309-0440}
\altaffiltext{4}{Johns Hopkins University, Baltimore, MD 21218}
\altaffiltext{5}{Harvard-Smithsonian Center for Astrophysics, Cambridge, MA 02138}
\altaffiltext{6}{Space Science and Technology Department, Rutherford Appleton Laboratory, Chilton, Didcot, Oxfordshire, OX11 0QX, U.K.}

\begin{abstract}
We present a survey of coronal forbidden lines detected in {\em Far Ultraviolet Spectroscopic Explorer} ({\em FUSE}) spectra of nearby stars.  Two strong coronal features, \ion{Fe}{18} $\lambda$974 and \ion{Fe}{19} $\lambda$1118, are observed in 10 of the 26 stars in our sample.  Various other coronal forbidden lines, observed in solar flares, also were sought but not detected.  The \ion{Fe}{18} feature, formed at $\log~T~({\rm K})~=~6.8$, appears to be free of blends, whereas the \ion{Fe}{19} line can be corrupted by a \ion{C}{1} multiplet.  {\em FUSE} observations of these forbidden iron lines at spectral resolution $\lambda/\Delta\lambda~\sim~15,000$ provides the opportunity to study dynamics of hot coronal plasmas.  We find that the velocity centroid of the \ion{Fe}{18} feature deviates little from the stellar rest frame, confirming that the hot coronal plasma is confined.  The observed line widths generally are consistent with thermal broadening at the high temperatures of formation and show little indication of additional turbulent broadening.  The fastest rotating stars, 31 Com, $\alpha$~Aur~Ab, and AB~Dor, show evidence for excess broadening beyond the thermal component and the photospheric $v\sin~i$.  The anomalously large widths in these fast rotating targets may be evidence for enhanced rotational broadening consistent with emission from coronal regions extending an additional $\Delta$R~$\sim$~0.4-1.3~R$_{\star}$ above the stellar photosphere or represent the turbulent broadening caused by flows along magnetic loop structures.  For the stars in which \ion{Fe}{18} is detected, there is an excellent correlation between the observed {\em R\"{o}ntgensatellit} ({\em ROSAT}) 0.2-2.0 keV soft X-ray flux and the coronal forbidden line flux.  As a result, \ion{Fe}{18} is a powerful new diagnostic of coronal thermal conditions and dynamics that can be utilized to study high temperature plasma processes in late-type stars.  In particular, {\em FUSE} provides the opportunity to obtain observations of important transition region lines in the far-UV, as well as, simultaneous measurements of soft X-ray coronal emission, using the \ion{Fe}{18} coronal forbidden line.  

\end{abstract}


\section{Introduction}

The study of stellar coronae historically has been relegated to X-ray wavelengths where high temperature plasmas predominately radiate.  Unfortunately, observational limitations including low spectral resolution, lack of precise wavelength calibrations, small effective areas, line blending, nonnegligible continua, and unreliable atomic parameters have made it difficult to transfer powerful spectroscopic techniques commonly used in the ultraviolet (UV) to the X-ray spectral region.  This disparity is improving with the development of powerful X-ray telescopes, such as the {\em Chandra X-ray Observatory} ({\em CXO}) and the {\em X-ray Multi-Mirror mission} ({\em XMM-Newton}).  With each new observation of a late-type star by these telescopes, our understanding of the structure of stellar coronae is advancing (e.g. Ayres et al. 2001; Brickhouse, Dupree, \& Young 2001; G\"{u}del et al. 2003).  However, even the modern array of X-ray spectrometers fall far short of the resolving power needed to study the dynamics of coronal emission lines.  In particular, the spectral resolution of the {\em CXO} gratings is nominally about $\lambda$/$\Delta\lambda~\sim~$1000, which corresponds to a velocity resolution of 300~km~s$^{-1}$.  The spectral resolution of the Reflection Grating Spectrometer (RGS) onboard {\em XMM-Newton} is at least a factor of two worse.  In contrast, the {\em Far Ultraviolet Spectroscopic Explorer} ({\em FUSE}) has a resolving power of $\lambda$/$\Delta\lambda~\sim~$15,000, which corresponds to a velocity resolution of 20~km~s$^{-1}$.  Presently, the fortuitous presence of coronal forbidden lines in the UV provides a way to study fundamental aspects of the structure of stellar coronae, by circumventing the limitations of contemporary X-ray spectroscopy.  

Over the course of the last century, coronal forbidden lines have repeatedly been observed and studied in the Sun \citep{edlen45}.  Most recently, solar flare observations using the {\em Solar Ultraviolet Measurements of Emitted Radiation} (SUMER) instrument onboard the {\em Solar \& Heliospheric Observatory} ({\em SOHO}) have identified many coronal forbidden lines in the wavelength range from 500-1600~\AA\ \citep{feldman00}.  By comparison, detections of ultraviolet coronal forbidden lines in other stars have been a relative novelty.  The initial discoveries were of \ion{Fe}{21} $\lambda$1354 in the brightest coronal late-type stars, including the active M-dwarf AU~Mic (M0~V; Maran et al. 1994), the hyperactive RS Canum Venaticorum (RS CVn) binary HR~1099 (K0~IV + G5~V; Robinson et al. 1996), and the giant-star binary system Capella (G8~III + G1~III; Linsky et al. 1998).  Unfortunately, the line profile analysis is complicated because \ion{Fe}{21} $\lambda$1354 is blended with a chromospheric \ion{C}{1} emission line.  Besides \ion{Fe}{21}, only one other coronal forbidden line species has been spotted in the wavelength range available from {\em Hubble Space Telescope} ({\em HST}): \citet{jordan01} identified the relatively cool \ion{Fe}{12} $\lambda\lambda$1242,1349 coronal lines in the spectrum of $\epsilon$~Eri.  \citet{ayres03} have presented a comprehensive survey of \ion{Fe}{12} and \ion{Fe}{21} coronal forbidden lines based on {\em HST} observations, primarily with the Space Telescope Imaging Spectrograph (STIS).

The {\em FUSE} satellite provides access to the important 900-1200~\AA\ spectral region below the short wavelength cutoff of {\em HST}, covering a new spectral region to search for stellar coronal forbidden lines.  Again, solar flare observations of coronal lines have previously been made at these wavelengths \citep{feldman00}.  \citet{feldman91} published the solar spectrum from 914-1177~\AA, based on rocket observations, as a template for stellar observations to be made by {\em FUSE}.  Included were 18 coronal forbidden lines, the strongest of which are: \ion{Ca}{13} $\lambda$1133, \ion{Fe}{18} $\lambda$974, \ion{Fe}{19} $\lambda$1118, and \ion{Fe}{17} $\lambda$1153.  Various authors have noted the presence of coronal forbidden lines in {\em FUSE} observations of active stars.  \citet{young01} identified the \ion{Fe}{18} line in the spectrum of Capella, and found that the bulk of the \ion{Fe}{18} emission arises from the G8 giant.  \citet{delzanna02} also used the \ion{Fe}{18} and \ion{Fe}{19} lines in their multi-wavelength analysis of the emission measure of AU~Mic.  \citet{redfield03} identified \ion{Fe}{18} and \ion{Fe}{19} in observations of AB~Dor, $\epsilon$~Eri, and AU~Mic, as part of a {\em FUSE} survey of cool dwarf stars.  \citet{dupree03} noted the \ion{Fe}{18} and \ion{Fe}{19} lines in $\beta$~Cet and 31~Com.

The present paper surveys a large body of {\em FUSE} observations of late-type stars for coronal forbidden lines.  Our objective is not only to enumerate positive detections, but also to demonstrate the utility of the coronal forbidden line diagnostics for understanding the dynamics of high temperature plasmas in stellar coronae.  

\section{Observations\label{secobs}}

Table~\ref{properties} lists the 26 stars in our sample.  They include the targets of the {\em FUSE} Team's cool star survey, supplemented with a number of Guest Observer observations.  Table~\ref{summary} summarizes the {\em FUSE} observations.
Most targets were observed in time-tagged mode, and all 
were observed either through the medium resolution slit (MDRS, $4.0^{\prime\prime} \times
20^{\prime\prime}$) or the large aperture (LWRS, $30^{\prime\prime} \times
30^{\prime\prime}$).  The LWRS has the advantage of usually including light from all of the LiF and SiC channels 
for the duration of the exposure, at the expense of higher airglow contamination, 
especially at Ly$\beta$ and Ly$\gamma$, but also in many \ion{O}{1} and \ion{N}{1}
lines.  The MDRS reduces airglow, but
thermal effects often cause one or more of the 
channels to drift out of the aperture during
an exposure, because only the LiF1 channel is used for guiding.  We monitor the integrated flux of the strongest lines in each segment (i.e. \ion{C}{3} $\lambda$977, \ion{O}{6} $\lambda$1031, and \ion{C}{3} $\lambda$1176).  During exposures when the star drifts our of the aperture, the flux in these lines is significantly reduced from its nominal value.  These exposures are therefore not included in the reduction of the final spectrum.

The spectral images were processed with CalFUSE 2.0.5\footnote{http://fuse.pha.jhu.edu/analysis/analysis.html}.  Compared to earlier versions of the pipeline, this new rendition provides a more accurate wavelength scale, removes
large noise event bursts, corrects for misplaced photons at low sensitivity portions of the
detector (so-called ``walk'' problem), improves 
corrections for scattered light and astigmatism, extracts the spectrum using an
``optimal'' algorithm, and applies a more accurate flux calibration (flux errors are $<$15\%; Oegerle, Murphy, \& Kriss [2000]\footnote{{\em FUSE} Data Handbook V1.1 is available at: http://fuse.pha.jhu.edu/analysis/dhbook.html.}).  Although the scattered light correction and flux calibration are much improved, they remain approximations and can occasionally lead to negative flux levels, where there is no stellar emission.  The flux error incorporates the errors involved in these calibration steps.  
Individual exposures were reduced separately and the integrated flux in strong emission lines was used to test the consistency of the signal recorded in each segment, throughout the entire observation.  For each segment, the exposures then were cross-correlated, using selected stellar emission features, and coadded, excepting those with significant drops in signal due to target drifts out of the aperture (a common occurrence for MDRS observations).  Redundant segments were not coadded, however, to avoid degrading the spectral resolution.  
The correction
for astigmatism (curvature of spectrum perpendicular to the dispersion)
improves the spectral resolution by 5--10\% compared with 
earlier versions of CalFUSE used to reduce the previously published spectra of AB~Dor \citep{ake00} and Capella \citep{young01}.

The relative wavelength scale within a given detector channel is
accurate to $\pm$~5~\kms\ over most of the far-ultraviolet (FUV) bandpass, except for some
deterioration at the far edges of the individual channels \citep{dixon01}\footnote{Introduction to CalFUSE v2.0 is available at: http://fuse.pha.jhu.edu/analysis/calfuse\_intro.html.}. Due to the absence of an onboard wavelength calibration lamp, we established 
the zero point of the wavelength scale according to interstellar absorption
features in the \ion{C}{2} 1036.34~\AA\ and \ion{C}{3} 977.02~\AA\ lines, 
and by comparing the \ion{C}{3} 1176~\AA\ lines in {\em FUSE} spectra with wavelength-calibrated observations 
by {\em HST}. The procedure is described in detail by \citet{redfield03}.

\section{Searching for Coronal Forbidden Lines\label{secsearch}}


We searched for evidence of coronal forbidden lines in the 900-1200~\AA\ region in our sample of stellar spectra.  Due to our relatively poor knowledge of the rest wavelengths of these lines, the relatively weak signal of the coronal lines, and the moderate likelihood of coincidental blends with low temperature lines, labeling spectral features located at wavelengths coincident with an expected coronal line can lead to misidentifications, as demonstrated below.  Therefore, we used a comparative technique developed by \citet{ayres03} to identify coronal forbidden lines in longer wavelength {\em HST} spectra.  We compared each sample spectrum with a typical spectrum of an active and inactive dwarf star (AU~Mic and $\alpha$~Cen~A, respectively), and an active and inactive giant star ($\beta$~Cet and $\beta$~Gem, respectively).  Strong coronal emission lines will stand out as broad features in the spectra of active stars, but should be absent in the inactive stars.  Coincidental low temperature lines in the active spectra can be identified because they will be relatively narrow, and also be present in the inactive stars.  The underlying assumptions in this comparative analysis are that active stars have larger coronal emission measures than inactive stars, and that giant stars all have reasonably similar emission measure distributions, and all dwarf stars also share similar emission measure distributions.  Although we have chosen specific active stars to demonstrate this technique below, any star with its own unique emission measure distribution can be used as the standard.  This comparison proves useful at identifying coincident blends and should be used along with other line identification techniques when searching for coronal forbidden lines.  

\begin{figure}
\epsscale{.69}
\figurenum{1}
\plotone{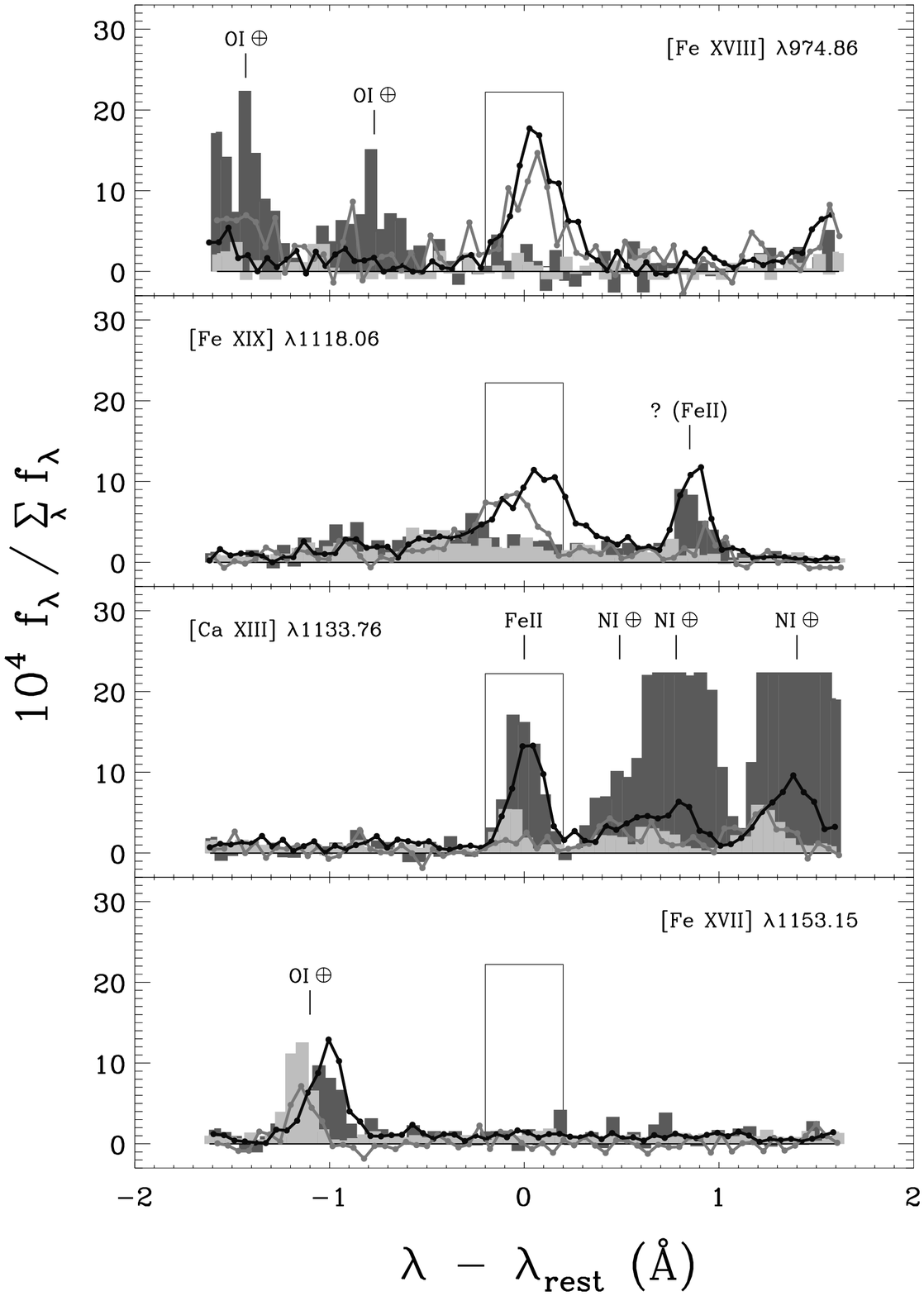}
\caption{Comparison technique to distinguish coronal emission lines from coincidental low temperature emissions.  Each of the four panels is the normalized stellar flux, centered on the rest wavelength of a strong coronal forbidden line in the {\em FUSE} spectral range, selected from solar work \citep{feldman91}.  Four different stars are shown: an active giant ($\beta$~Cet {\it dark connected circles}), an inactive giant ($\beta$~Gem {\it dark filled histogram}), an active dwarf (AU~Mic {\it light connected circles}), and an inactive dwarf ($\alpha$~Cen~A {\it light filled histogram}).  The central box indicates the wavelength region where the coronal feature is expected to be found.  Genuine coronal forbidden lines should appear in the spectra of the active stars, but not the inactive ones.  The top two panels show positive detections of \ion{Fe}{18} $\lambda$974 and \ion{Fe}{19} $\lambda$1118.  The third panel demonstrates that the comparative technique successfully ferrets out coincidental blends.  For example, there appears to be a feature that corresponds to \ion{Ca}{13} $\lambda$1133, but is detected in the active and inactive giants only.  The feature, in fact, is a blend of two fluorescent \ion{Fe}{2} lines at 1133.68~\AA\ and 1133.85~\AA\ \citep{harper01}.  The bottom panel demonstrates that \ion{Fe}{17} $\lambda$1153 is absent, at {\em FUSE} sensitivity levels.  \label{lines}}
\end{figure}

Figure~\ref{lines} illustrates the comparative technique.  We selected the four strongest high temperature ($T~>~10^6$~K) coronal forbidden lines observed in solar flares \citep{feldman91} and located in the {\em FUSE} spectral range: \ion{Fe}{18} $\lambda$974 ($\log T~\sim~6.81$~K), \ion{Fe}{19} $\lambda$1118 ($\log T~\sim~6.89$~K), \ion{Ca}{13} $\lambda$1133 ($\log T~\sim~6.38$~K), and \ion{Fe}{17} $\lambda$1153 ($\log T~\sim~6.72$~K).  As the strongest lines in the {\em FUSE} spectral range detected in solar flares, these lines are an appropriate place to start looking for coronal forbidden lines in stellar spectra.  The top two panels of Figure~\ref{lines} demonstrate that \ion{Fe}{18} and \ion{Fe}{19} are positively detected, because the features appear in the normalized spectra of both active stars, and are absent in the inactive stars.  \ion{Fe}{17}, shown in the bottom panel is clearly not detected in any of the stars.  At the expected location of \ion{Ca}{13}, there is a prominent feature.  However, it is present only in the spectra of the giant stars, both active and inactive, and is narrow, inconsistent with the high line formation temperature.  Therefore, it has the characteristics of a low temperature line.  In fact, \citet{harper01} identified numerous fluorescent \ion{Fe}{2} lines in the {\em FUSE} spectra of $\alpha$~TrA, two of which fall at wavelengths 1133.675~\AA\ and 1133.854~\AA.  These lines are blended and appear as a single feature.  Therefore, the emission detected in {\em FUSE} spectra at $\sim~1133.7$~\AA\ is not \ion{Ca}{13}, but rather a coincidental fluorescent \ion{Fe}{2} line pumped by H~Ly$\alpha$.

The comparative technique was applied to an extensive list of more than 50 candidate coronal forbidden lines that lie in the {\em FUSE} spectral range.  Of these, only two were positively detected: the \ion{Fe}{18} $\lambda$974 and \ion{Fe}{19} $\lambda$1118 lines mentioned previously.  These features form at $\log~T~({\rm K})~=~6.8$ and $6.9$, respectively.  In Table~\ref{observed}, we list the observed integrated line strengths corresponding to direct flux integrations within $\pm$~150~km~s$^{-1}$ of line center for stars with $v\sin i~<~50$~km~s$^{-1}$, and within $\pm$~250~km~s$^{-1}$ for stars with $v\sin i~>~50$~km~s$^{-1}$.  X-ray fluxes, also included in the table, are based on the 0.2-2.0~keV band observed by the {\em R\"{o}ntgensatellit} ({\em ROSAT}).  
 
\section{\ion{Fe}{18} $\lambda$974\label{secfe18}}
  
The addition of the \ion{Fe}{18} $\lambda$974 line, in particular, to the collection of UV coronal forbidden lines is very promising, due to its optimum spectral context.  Unlike other high temperature forbidden lines, such as \ion{Fe}{19} $\lambda$1118 and \ion{Fe}{21} $\lambda$1354 whose profiles are compromised by coincident blends\footnote{The \ion{Fe}{21} line is blended with the \ion{C}{1} line at 1354.29~\AA, and \ion{Fe}{19} is blended with eight \ion{C}{1} lines from 1117.20~\AA\ to 1118.49~\AA.}, the nearest significant blend with \ion{Fe}{18} is $\sim~300$~km~s$^{-1}$ from line center.  Therefore, \ion{Fe}{18} requires no additional assumptions or delicate deconvolutions, when trying to reconstruct the intrinsic coronal forbidden line profile.  Furthermore, the ultraviolet continua of typical cool stars decrease dramatically toward the shorter wavelengths.  Therefore, modeling the continuum, if present at all, is less challenging in the spectral region around \ion{Fe}{18}, than, for example, at \ion{Fe}{21} $\lambda$1354.  Also, the strong local interstellar medium (LISM) absorption in the prominent \ion{C}{3} 977.02~\AA\ line, adjacent to the \ion{Fe}{18} line, provides a convenient means of calibrating the zero point of the local velocity scale.  

\begin{figure}
\epsscale{.85}
\figurenum{2}
\plotone{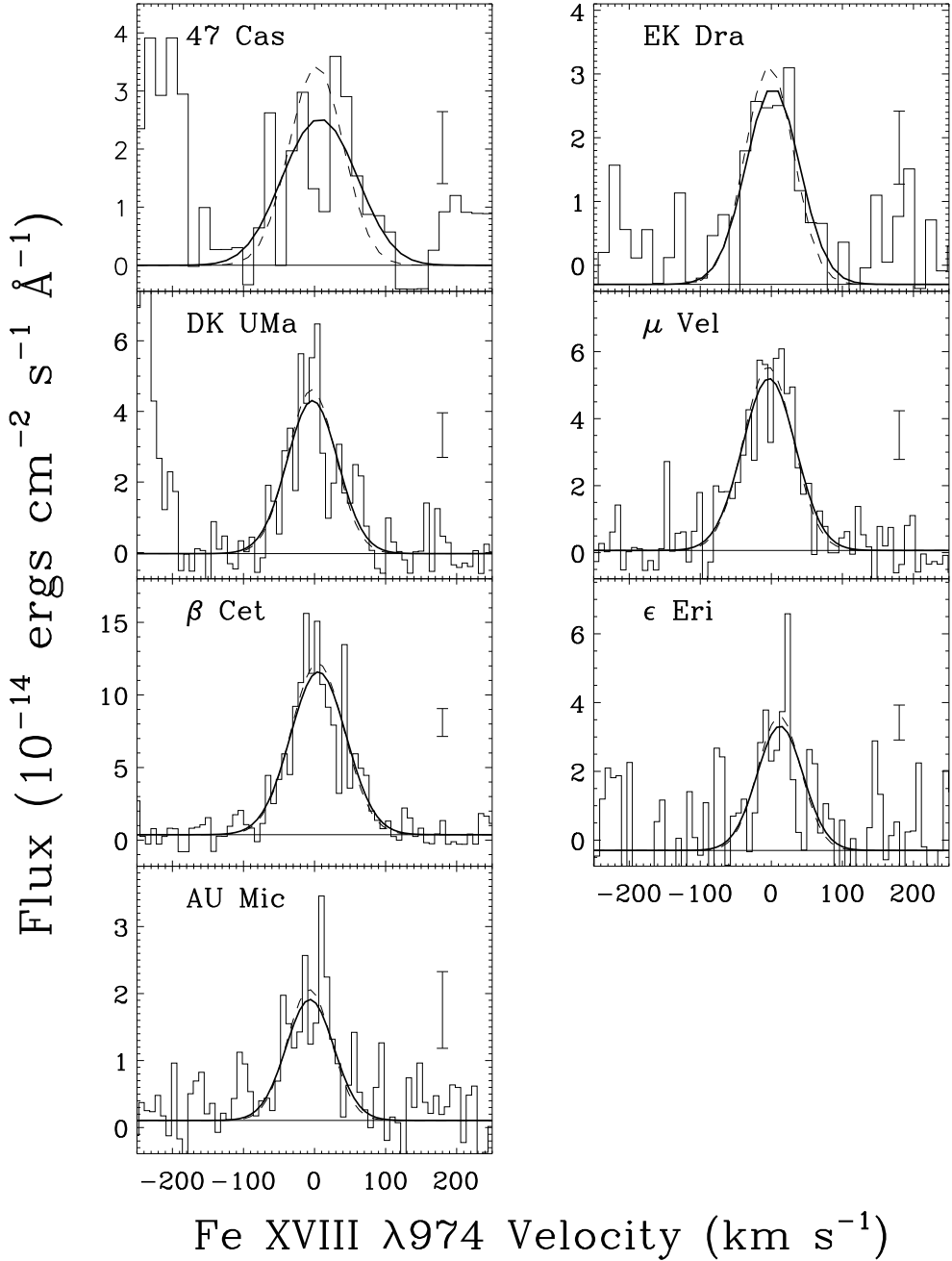}
\caption{Single Gaussian fits to \ion{Fe}{18} $\lambda$974 for the slowly rotating stars.  The SiC~2A channel data are shown as a histogram, and zero velocity is fixed at the stellar (photospheric) radial velocity.  The continuum (horizontal solid line) is assumed to be constant over the line profile and is matched to line-free regions on either side of the coronal feature.  The dashed curve indicates the intrinsic line profile before convolution with the rotational and instrumental line profiles, and the solid curve is the result after the broadening functions have been applied.  A 1~$\sigma$ error bar is shown on the right side of each panel.  \label{fe18line}}
\end{figure}

Figure~\ref{fe18line} presents the \ion{Fe}{18} $\lambda$974 lines, and single Gaussian fits, in seven spectra of the slowly rotating stars with \ion{Fe}{18} detections.  Table~\ref{fits} lists the derived parameters.  In Figure~\ref{fe18line}, the solid horizontal line indicates the continuum level (often zero), based on line-free regions on either side of the \ion{Fe}{18} feature.  The intrinsic stellar line profile (prior to convolution with the instrumental line spread function [Wood et al. 2002] and the rotational line profile) is shown by the dashed curve.  The solid curve represents the predicted profile including both instrumental and rotational broadening.  Note that the intrinsic line widths are so broad that the contribution of the instrument and rotation often is negligible.  Rotational broadening is most noticeable for 47~Cas, which is the only star among these seven with $v\sin~i~>~20$~km~s$^{-1}$; the other six stars have relatively modest rotational velocities (see Table~\ref{properties}).  In fact, Figure~\ref{fe18line} indicates very little variation in line width among the seven targets, consistent with domination by thermal broadening.  The line formation temperature for \ion{Fe}{18} is $\log~T~({\rm K}) = 6.76$, which corresponds to a full-width at half-maximum (FWHM) of 69~km~s$^{-1}$, but \ion{Fe}{18} is likely to form over a modest spread in temperatures.  A reasonable range is $6.62 \leq \log~T~({\rm K}) \leq 6.93$, which corresponds to widths of $59 \leq {\rm FWHM}$~km~s~$^{-1} \leq 84$.  In column~7 of Table~\ref{fits}, we list the equivalent temperature required to account for the observed intrinsic line width if turbulent motions do not contribute to the line broadening.    

Figure~\ref{fe18line} also demonstrates that \ion{Fe}{18} $\lambda$974 displays no large velocity shifts, as the emission centroids are approximately at the rest velocity of the stellar photosphere in all cases.  In a couple of the stars, 47~Cas and EK~Dra, we were unable to determine an accurate zero point of the wavelength scale because either there is no obvious LISM absorption in the \ion{C}{3} 977.02~\AA\ line, or no high-resolution wavelength-calibrated data (e.g., STIS) available to measure the velocity of LISM absorption empirically.  Therefore, for these two stars, we assume that the centroid of the \ion{C}{3} feature is at the rest velocity of the star.  \citet{redfield03} showed that the centroid of \ion{C}{3} can be shifted from line center by as much as 15~km~s$^{-1}$.  Therefore we use 15~km~s$^{-1}$ as our estimate of the systematic error.  Only in the two broadest profiles, 31~Com and AB~Dor, do the statistical errors dominate the systematic errors.  Spreading out the signal of an already weak coronal line, makes it extremely difficult to accurately determine the centroid of the \ion{Fe}{18} feature.  Higher S/N observations are required for the stars that exhibit broad \ion{Fe}{18} line profiles in order to measure the centroids of the emission.  

\begin{figure}
\epsscale{.7}
\figurenum{3}
\plotone{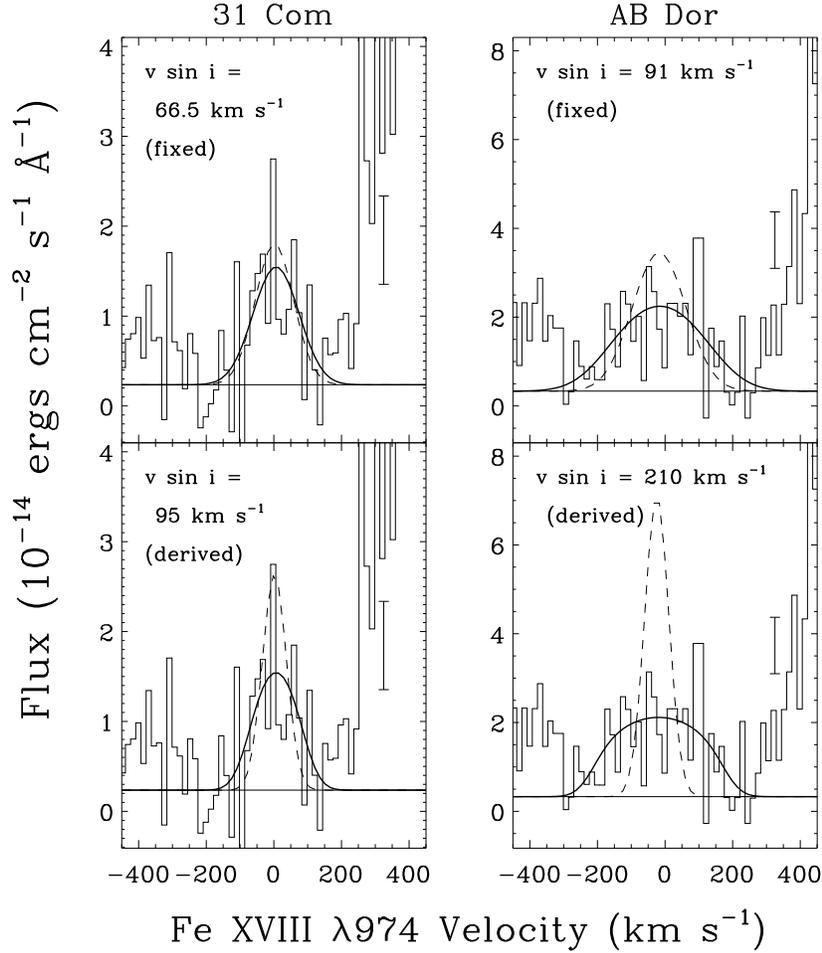}
\caption{Single Gaussian fits to \ion{Fe}{18} $\lambda$974 for the fast rotating stars, 31~Com and AB~Dor.  The data shown are taken from the SiC~2A channel.  The continuum (horizontal solid line) is assumed to be constant over the line profile and is matched to line-free regions in the vicinity of the coronal feature.  The top panels show the result when the $v\sin i$ is fixed at the photospheric value.  The sharp rise in flux to the red of the coronal line is the blue wing of the strong \ion{C}{3} $\lambda$977 feature.  The inferred intrinsic line widths including rotation (dashed curves) are uncharacteristically large in both cases.  The solid lines show the convolution of the intrinsic profiles with the instrumental profiles. The bottom panels show an alternative model in which the intrinsic line width (dashed curve) is frozen at the thermal value and the rotational velocity ($v\sin~i$) is allowed to vary.  The best fit $v\sin i$ is given in the bottom panels.  The ``super-rotational'' broadening can be explained by highly extended coronal regions co-rotating with the stellar surface.  A 1~$\sigma$ error bar is shown on the right side of each panel.  \label{fe18rot}}
\end{figure}

The \ion{Fe}{18} widths in most of the stars are consistent with little to no turbulent broadening: Thermal broadening, alone, is sufficient to account for the observed intrinsic line width.  The two fastest rotating stars in the sample, 31~Com and AB~Dor show large intrinsic line widths, however, even when the photospheric rotational broadening is taken into account.  
Another fast rotator, 47~Cas, also shows a large line width, marginally in excess of the expected thermal broadening.  Although an additional broadening mechanism may also play a part in the coronal line profiles of 47~Cas, we limit the following analysis to 31~Com and AB~Dor.  The spectra of these two stars clearly indicates that thermal broadening alone is insufficient to account for the width of the coronal line profiles.
In Figure~\ref{fe18rot}, the top panels show the initial single Gaussian fit for which the rotational broadening was fixed at the observed photospheric value.  It is clear that the inferred intrinsic line profile is much wider than those of the slow rotators, shown in Figure~\ref{fe18line}.  It seems unlikely that nonthermal, turbulent broadening in 31~Com and AB~Dor could explain the very broad profiles, because such broadening is clearly of little importance in the other stars.  Instead, we interpret the additional broadening to be the result of ``super-rotation'', extended co-rotating coronal emission high above the stellar photosphere.  In the bottom panels of Figure~\ref{fe18rot}, the intrinsic line width was fixed at the thermal value.  An alternative broadening mechanism is now required to derive a satisfactory fit to the data.  If we allow the rotational velocity to vary, the additional broadening can now be attributed to a value of $v\sin~i$ that is larger than the photospheric value.  Both models produce equally satisfactory fits, but the top panels indicate a model that assumes that the rotational velocity is fixed at the photospheric value and turbulent broadening must account for the line width, whereas the bottom panels assume that turbulent broadening is negligible, based on the results of the slowly rotating stars, and the rotational velocity is instead the free parameter.  For 31~Com, $\chi^2_{\nu} = 1.25$ for the fixed rotational velocity model and $\chi^2_{\nu} = 1.24$ for the variable rotational velocity model.  For AB~Dor, $\chi^2_{\nu} = 2.48$ for the fixed rotational velocity model and $\chi^2_{\nu} = 2.42$ for the variable rotational velocity model.  Unfortunately, the S/N of the data is not high enough to differentiate between the two assumptions based solely on the subtle differences in the resulting line profile.  However, the unanimous result that turbulent broadening is nonexistent in our sample of slowly rotating stars, leads us to propose that the additional broadening is due to the rotation of the coronal material at velocities greater than the photospheric $v\sin~i$.  The best fit $v\sin~i$ for 31~Com is $95^{+80}_{-20}$ km~s$^{-1}$ and for AB~Dor, $210^{+60}_{-40}$ km~s$^{-1}$.  Although the S/N of the data prohibits an accurate determination of the coronal rotational velocity, it is clear that it must be greater than the photospheric value.  Based on the enhanced rotational velocities, we can estimate how far above the stellar surface the presumably co-rotating emitting plasma extends.  For 31~Com, we infer that coronal material extends to $\Delta$R~$\sim~0.4^{+1.2}_{-0.3}$~R$_{\star}$ above the stellar surface, and for AB~Dor, $\Delta$R~$\sim~1.3^{+0.7}_{-0.4}$~R$_{\star}$.  Similar calculations have been made using transition region lines \citep{ayres98,brandt01} and coronal forbidden lines \citep{ayres03}, observed with {\em HST}.  Extended, but cooler, structures also have been inferred through time series of H$\alpha$ absorption spectra of AB~Dor \citep{cameron96}.  \citet{jardine02} have used Zeeman-Doppler maps of AB~Dor to extrapolate the topology of the extended coronal magnetic field.

An alternative interpretation to explain the observed additional broadening is by the presence of high speed flows.  Measurements of density sensitive lines in spectra observed by the {\em Extreme Ultraviolet Explorer} satellite ({\em EUVE}), seem to indicate that the emitting volume of the coronal plasma is small compared to the stellar radius \citep{dupree93,sanz02}.  This appears to imply compact coronal structures as opposed to extended co-rotating coronal emission.  Such high speed flows in coronal plasmas have recently been observed on the Sun \citep{winebarger02}.  However, we only detect additional broadening in the fastest rotating stars, not necessarily in the most active.  For example, slow rotators, such as $\epsilon$~Eri and AU~Mic, which are much more active than the Sun, do not exhibit additional broadening, although they may be expected to have prominent high speed flows.  Therefore, we feel that emission from plasma confined in extended but thin magnetic loops is a more likely explanation for the observed additional broadening in the fastest rotators, although a specific model should be built with densities consistent with the EUVE measurements to support this conclusion.

To estimate whether plasma confined in extended magnetic flux tubes can explain the observed broad Fe~XVIII emission lines, we consider this simple example. We assume that all of the Fe~XVIII emission occurs in half-circular flux loops extending up to a height $\Delta R$ above the stellar surface with a constant radius $\alpha \Delta R$ and electron density $n_e$. If there are $n$ such loops, then the emission measure $EM = n_e^2 n V_{loop} =  n_e^2 n \pi^2 \alpha^2 (\Delta R)^3$. We can now solve for the product, 

\[ n\alpha^2 = \frac{EM}{\pi^2 n_e^2 (\Delta R)^3}.  \]

As an example, we consider the corona of AB~Dor. \citet{gudel01} analyzed the nonflare spectrum of AB~Dor observed by the RGS instrument on XMM. In their three component model, the middle temperature component has a temperature $kT_2 = 0.65$ keV, similar to the formation temperture of Fe~XVIII, and the corresponding emission measure $EM_2 = 4.7\times 10^{52}$ cm$^{-3}$. Ratios of the helium-like triplet lines of O~VII are consistent with a coronal electron density $n_e = 3\times 10^{10}$ cm$^{-3}$. If ``super-rotation'' is the explanation for the widths of the Fe~XVIII line, then $\Delta R = 1.3 R_{\star} = 9 \times 10^{10}$ cm for $R_{\star} \sim 1 R_{\odot}$ \citep{gudel01}. These parameters yield $n \alpha^2 = 0.007$. If we assume n = 10 or 100, then $\alpha$ = 0.03 or 0.009, and the loops are thin as is true for the solar corona \citep{schrijver99}. We conclude that the ``super-rotation'' model with many thin flux loops is consistent with the coronal data for this test case. Also the flux loop thickness parameter $\alpha$ will likely be small, as expected, for a wide range of coronal parameters.

\subsection{\ion{Fe}{18} $\lambda$974 in Capella\label{secfe18cap}}

Interpretation of the coronal \ion{Fe}{18} $\lambda$974 feature of the giant binary Capella is complicated by the fact that both stars (the G8 giant primary [$\alpha$ Aur Aa] and the G1 giant secondary [$\alpha$ Aur Ab]) can contribute to the \ion{Fe}{18} emission.  {\em HST} observations of the \ion{Fe}{21} $\lambda$1354 line profile of Capella have demonstrated that the relative contribution by each star changes with time.  Based on the analysis of Capella observations taken on 1995 September 9, \citet{linsky98} determined that each star provided approximately half of the total \ion{Fe}{21} emission flux.  \citet{johnson02} and \citet{ayres03} showed in STIS observations of Capella taken four years later, on 1999 September 12, that the \ion{Fe}{21} emission from the G8 primary star was reduced to $\lesssim~20$\% of its previous level, although the G1 component remained about the same.  \citet{young01} used a single Gaussian model and applied it to the first {\em FUSE} observation of Capella (2000 Nov 5) to evaluate the individual stellar contributions to the \ion{Fe}{18} emission flux.  They estimated that the G8 primary contributed $\sim~75$\% of the total flux.  We go beyond that first stage analysis by fitting all three available {\em FUSE} observations in an effort to better understand the nature of Capella's coronal \ion{Fe}{18} feature.  The dates of the Capella observations are listed in Table~\ref{summary}, while individual stellar velocities, phases, and the ephemeris reference are listed in Table~\ref{properties}.  

\begin{figure}
\epsscale{.85}
\figurenum{4}
\plotone{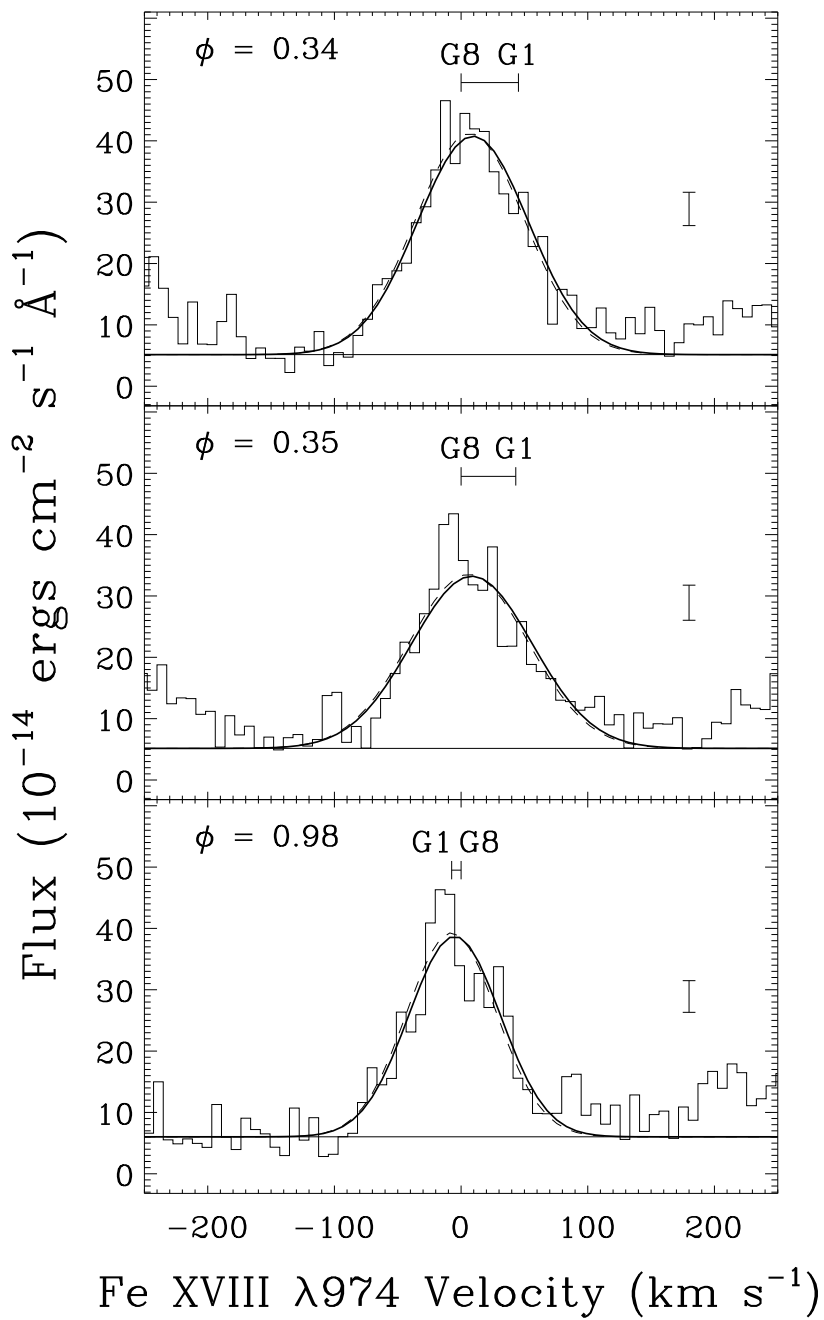}
\caption{Single Gaussian fits to \ion{Fe}{18} $\lambda$974 for Capella, observed at three orbital phases:  The first two observations were taken near quadrature, the third near conjunction.  The velocity scale is in the reference frame of the G8~III star.  The relative velocity of the G1~III star also is indicated.  The data shown are taken from the SiC~2A channel.  A 1~$\sigma$ error bar is shown on the right side of each panel.  \label{fe18cap}}
\end{figure}

First, we modeled the \ion{Fe}{18} $\lambda$974 emission profile with a single Gaussian, appropriate if the emission were coming principally from one of the stars.  Figure~\ref{fe18cap} illustrates the fit to each individual exposure.  The first two observations were taken near orbital quadrature (maximum radial velocity separation), and the third near conjunction (minimum velocity separation).  The velocity scale was adjusted to the reference frame of the G8 primary star, known to $\leq~5$~km~s$^{-1}$, using the strong LISM absorption in the neighboring \ion{C}{3} 977.02~\AA\ line \citep{wood03}.  The fit parameters corresponding to Figure~\ref{fe18cap} are listed in Table~\ref{fitscapella}.  It is unclear why there is a decrease in integrated flux between the first and second two observations.  Instrumental effects, such as thermal drifts, can cause a decrease in flux, as discussed in Section~2.  Although stellar variability may also explain the decrease in flux \citep{ambruster87}, it is unlikely because the first two discrepant observations were are taken only two days apart.  However, the flux discrepancy does not effect the line width, line centroid, or flux ratio of the binary components, and therefore does not effect the conclusions of this work.  The bulk of the emission clearly is associated with the G8 star, based on the centroid of the emission, as was noted by \citet{young01}.  Note, however, that the \ion{Fe}{18} line width decreases significantly between quadrature and conjunction.  This indicates that although emission from the G8 primary star may dominate, there is a nonnegligible contribution from the faster rotating G1 star, and we are compelled to fit the Capella emission feature with two components, one for the contribution of each star.

\begin{figure}
\epsscale{.85}
\figurenum{5}
\plotone{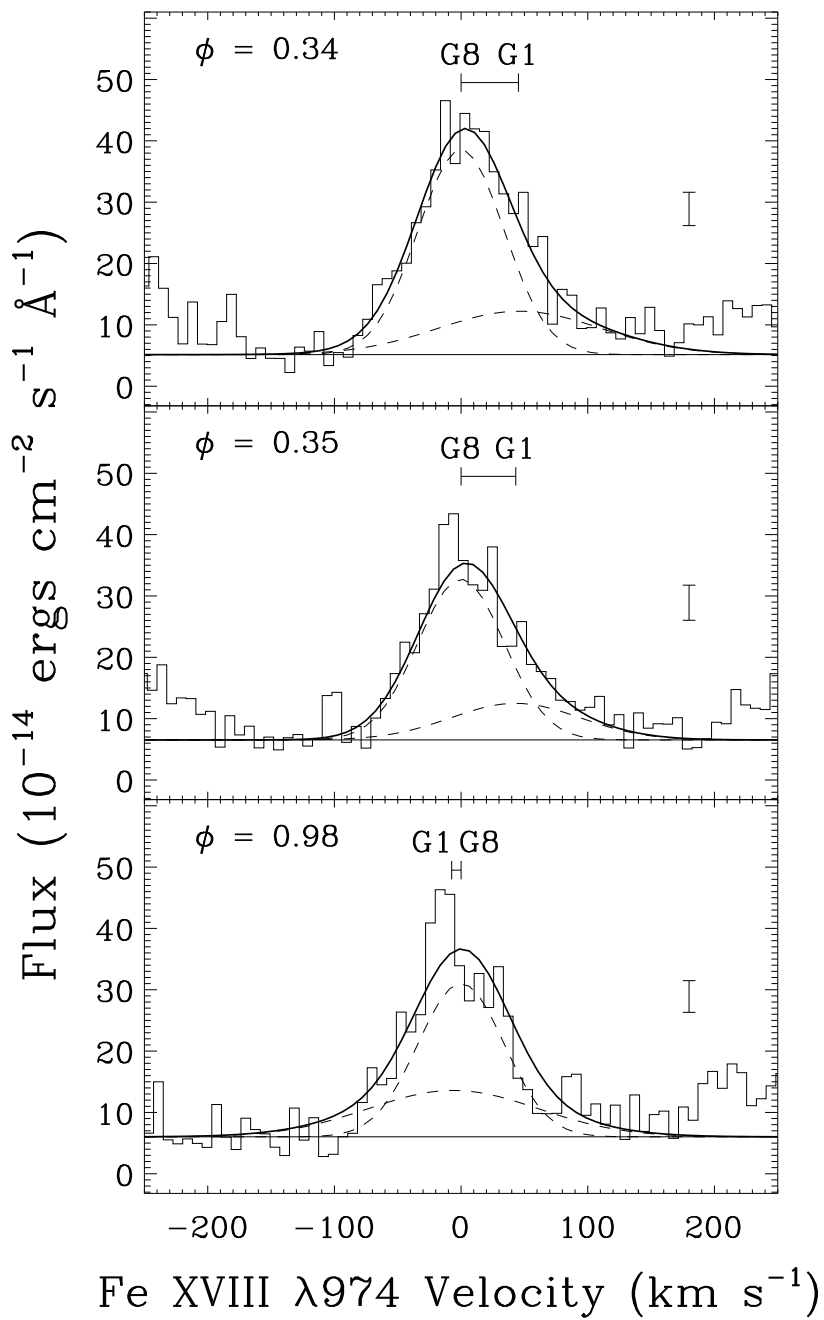}
\caption{Same as Figure~\ref{fe18cap}, but for a multiple Gaussian model, with one component for each star.  The velocity centroids of both stars, and the FHWM of the primary were fixed in the quadrature spectra.  The only free parameter in the model for the conjunction spectrum is the total flux.  A 1~$\sigma$ error bar is shown on the right side of each panel.  \label{fe18_2cap}}
\end{figure}

In order to estimate the contribution of the G1 secondary star, we modeled the \ion{Fe}{18} $\lambda$974 emission profile with two Gaussians, one for each star, including instrumental and rotational broadening as before.  In this case, however, the rotational velocity of the G8 star is small ($v\sin~i~=~3~$km~s$^{-1}$), but that of the G1 star is much larger ($v\sin~i = 36$~km~s$^{-1}$; Strassmeier et al. [1988]).  Although there is discernible emission from the G1 secondary star, it is still relatively weak, and the Gaussian modeling benefits from fixing several of the free parameters.  In particular, we constrained the velocity centroid of each component to be at the respective orbital velocity, and the FWHM of the G8 component to the value determined from the single Gaussian fit near conjunction (essentially the thermal width).  Both constraints are motivated by the empirical \ion{Fe}{18} behavior of active single stars, discussed in Section~\ref{secfe18}: Velocity centroids are consistent with the photospheric radial velocities; and line widths in slow rotators are well characterized by thermal broadening alone.  The results of the double Gaussian modeling are depicted in Figure~\ref{fe18_2cap}, and listed in Table~\ref{fitscapella}.  The $\chi^2_{\nu}$ goodness-of-fit metric is also listed in Table~\ref{fitscapella}.  The fits are noticeably improved for the two observations near quadrature, particularly redward of the main component (80-140 km~s$^{-1}$).  The G1 secondary apparently contributes $\sim~30$\% of the flux in \ion{Fe}{18}.  The width of the secondary component is broader than expected from thermal broadening alone.  As for 31~Com and AB~Dor discussed previously, we interpret the excess width in terms of an extended coronal zone.  In this case, we estimate that the emission extends $\Delta$R~$\sim~1.2~$R$_{\star}$ above the stellar photosphere.  \citet{ayres03} reached similar conclusions in the \ion{Fe}{21} $\lambda$1354 line of the G1 star.

\section{\ion{Fe}{19} $\lambda$1118\label{secfe19}}

\begin{figure}
\epsscale{.85}
\figurenum{6}
\plotone{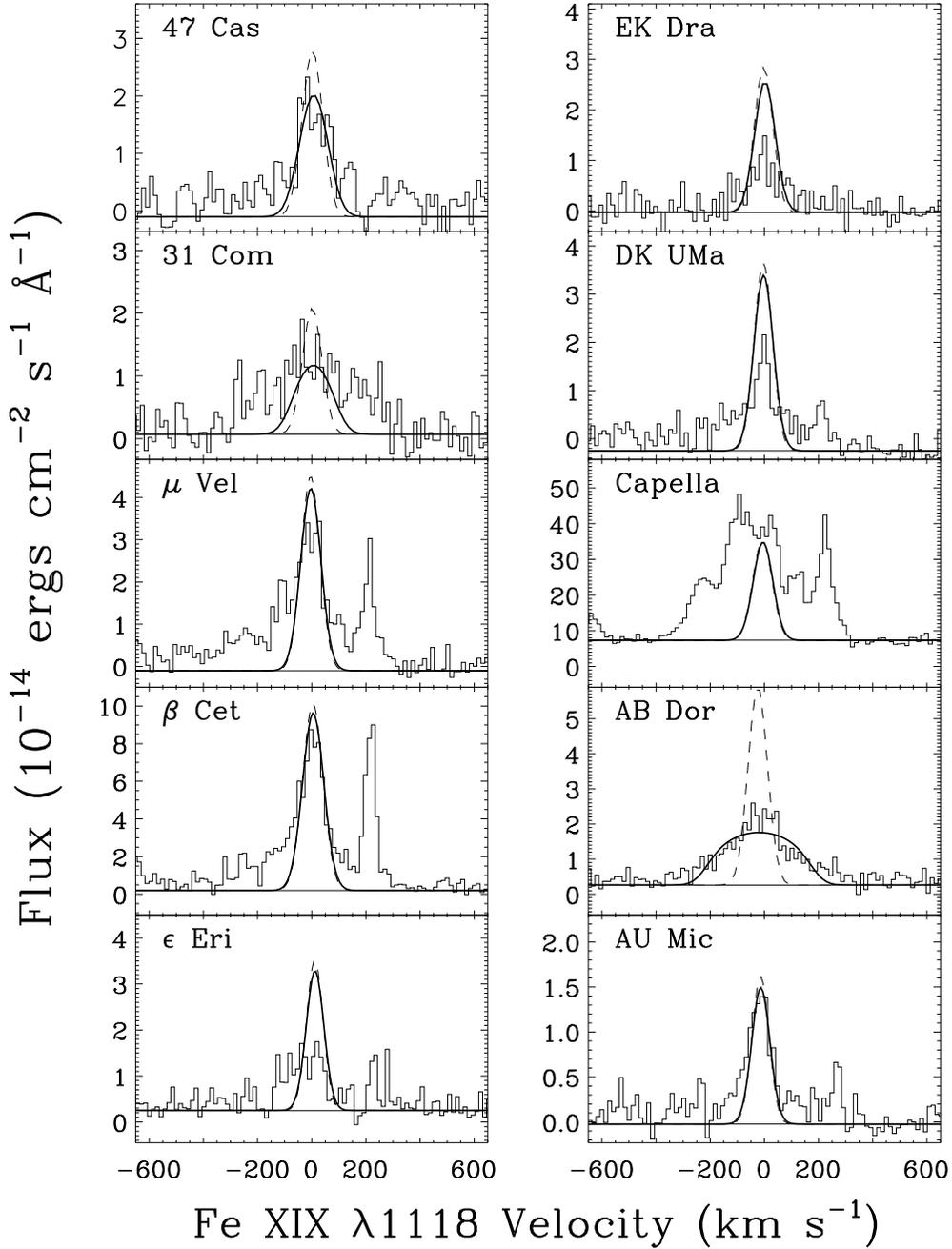}
\caption{Predicted emission profiles of \ion{Fe}{19} $\lambda$1118.  The LiF~2A channel data are shown as a histogram, where zero velocity is fixed at the stellar radial velocity.  The continuum level is indicated by the horizontal solid line.  The \ion{Fe}{18} profile fits in Figures~\ref{fe18line}-\ref{fe18cap} have been used to predict the \ion{Fe}{19} emission profiles.  Excess emission is evident in some cases owing to a severe blend with a \ion{C}{1} multiplet.  \label{fe19line}}
\end{figure}

The other coronal forbidden line detected in {\em FUSE} observations of late-type stars is \ion{Fe}{19} $\lambda$1118.  In contrast to \ion{Fe}{18} $\lambda$974, there are several severe blends that can obscure the \ion{Fe}{19} feature.  At least eight \ion{C}{1} lines are scattered between 1117.20~\AA\ and 1118.49~\AA.  In some stars, however, these \ion{C}{1} lines are weak relative to the coronal emission.  The interval surrounding \ion{Fe}{19} is shown in Figure~\ref{fe19line} for the 10 stars that exhibit a significant \ion{Fe}{18} feature.  If we assume that the emission measure is the same for the formation of \ion{Fe}{18} and \ion{Fe}{19}, we can predict the line profile of \ion{Fe}{19} based on theoretical emissivities and the empirical Gaussian parameters.  We compute the emissivities with CHIANTI version 4 \citep{dere97,young02}.  Each ion is assumed to be emitting at its most probable temperature.  The resulting ratio of emissivities is insensitive to assumptions of pressure, abundances, or ionization equilibrium.  In Figure~\ref{fe19line}, the single Gaussian profiles of \ion{Fe}{18}, presented in Figures~\ref{fe18line}, \ref{fe18rot}, and \ref{fe18cap}, have been scaled by the relative emissivities.  In some cases, particularly 47~Cas, $\beta$~Cet, AU~Mic, and $\mu$~Vel, there is excellent agreement between the observed spectra and the predicted \ion{Fe}{19} line profile.  The shapes and Doppler shifts of the \ion{Fe}{19} line profiles relatively free of blending, seem to be identical to \ion{Fe}{18}.  Here the \ion{C}{1} lines are weak, although some excess carbon emission is present, particularly blueward of the coronal feature.  In other cases, especially Capella and 31~Com, the \ion{C}{1} emission wreaks havoc with the spectra.  The \ion{Fe}{19} line is clearly detected in many cases, nevertheless, severe blending typically limits our ability to analyze it.  In some cases, particularly, EK~Dra, DK~UMa, and $\epsilon$~Eri, we significantly overestimate the flux of the \ion{Fe}{19} feature.  Our first order assumption that the emission measure is identical for \ion{Fe}{18} ($\log T~\sim~6.81$~K) and \ion{Fe}{19} ($\log T~\sim~6.89$~K) was motivated by the idea that the emission measure distribution is unlikely to change drastically over a small change in formation temperature.  This is supported by the differential emission measure analysis of AU~Mic by \citet{delzanna02}, as well as the other stars that show excellent agreement between the observed and predicted levels of \ion{Fe}{19} emission in Figure~\ref{fe19line}.  

Because stellar flux was not detected in the SiC channels of the 1999 October 20 observation of AB~Dor, the \ion{Fe}{18} emission profile was only observed in the 1999 December 14 observation (see Table~\ref{summary}).  However, both visits successfully observed the spectral region surrounding the \ion{Fe}{19} coronal forbidden line.  The coadded spectrum is shown in Figure~\ref{fe19line}.  The full {\em FUSE} spectrum of AB~Dor indicates that \ion{C}{1} emission lines are present \citep{redfield03}.  Therefore, we expect some \ion{C}{1} contamination in the \ion{Fe}{19} profile.  The broad emission in the \ion{Fe}{19} line of AB~Dor appears to be consistent with the broad emission detected in \ion{Fe}{18}, although it is difficult to differentiate between the contribution of \ion{Fe}{19} flux and \ion{C}{1} emission.  This is an excellent example of the powerful use of a coronal forbidden line that is uncontaminated by low temperature lines, such as \ion{Fe}{18} $\lambda$974, and the difficulty in analyzing coronal lines that are severely blended, such as \ion{Fe}{19} $\lambda$1118.

\section{\ion{Fe}{18} and Soft X-rays\label{secxray}}

\begin{figure}
\epsscale{.7}
\figurenum{7}
\plotone{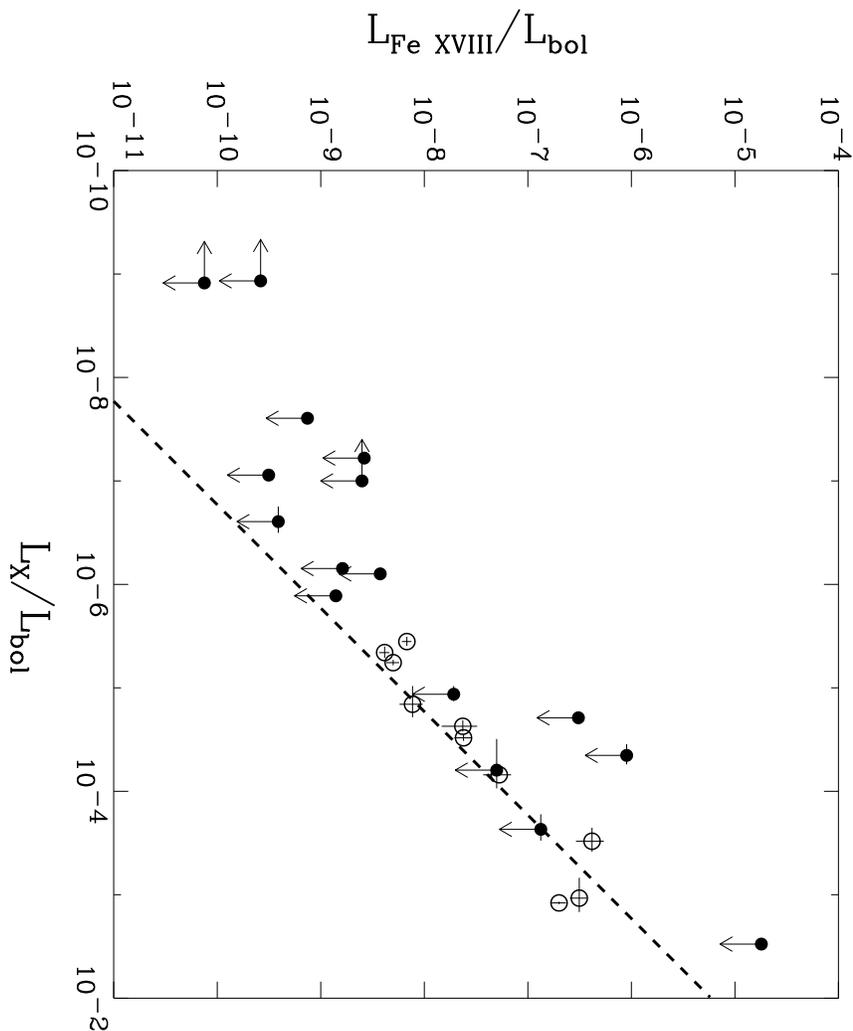}
\caption{Comparison of normalized \ion{Fe}{18} $\lambda$974 fluxes and soft X-rays.  Large open circles indicate stars with detected \ion{Fe}{18} emission; small filled symbols indicate upper limits.  The detections are well matched by a unit slope power law correlation (thick dashed line).  An anticipated break at lower activity levels is masked by the apparent sensitivity levels of the typical {\em FUSE} pointings. \label{lfelx}}
\end{figure}

All of the stars in our sample have soft X-ray fluxes observed by {\em ROSAT}.  Because the forbidden lines observed in the UV and the soft X-rays both sample high temperature coronal plasmas in these stars, a correlation between the two diagnostics is expected.  Figure~\ref{lfelx} compares \ion{Fe}{18} $\lambda$974 and the 0.2-2.0 keV soft X-rays.  The coronal fluxes are normalized by the bolometric fluxes given in Table~\ref{properties}, to remove biases of stellar size and distance.  Those stars with detected \ion{Fe}{18} emission are displayed as open circles, and those with upper limits are shown as small filled circles.  A unit slope power law is indicated by the thick dashed line.  A good correlation exists over two orders of magnitude, down to a sensitivity threshold of $L_{\rm Fe~XVIII}/L_{\rm bol}~\sim~3~\times~10^{-9}$.  All of the upper limits are consistent with this correlation.  \citet{ayres03} demonstrated a similar unit power law correlation for \ion{Fe}{21} $\lambda$1354.  In active stars, at least, the \ion{Fe}{18} flux can serve as a surrogate for the broad-band soft X-ray emission.

\section{Conclusions\label{secconc}}

{\em FUSE} provides an opportunity to study a new collection of high temperature ($T > 10^6$ K) forbidden lines in nearby coronal stars.  The goal of the present paper was to extensively examine {\em FUSE} observations of late-type stars for such features, and to analyze any detections as potential diagnostics of the physical characteristics of the high temperature plasmas.  Our results are as follows:

\begin{itemize}

\item  After searching the {\em FUSE} spectra for candidate coronal forbidden lines identified in solar flares \citep{feldman91,feldman00}, only two were found: \ion{Fe}{18} $\lambda$974 and \ion{Fe}{19} $\lambda$1118.  Detections were obtained in 10 of the 26 stars in our sample.  The two {\em FUSE} lines join \ion{Fe}{12} $\lambda\lambda$1242,1349 and \ion{Fe}{21} $\lambda$1354 coronal lines from the STIS 1200-1600~\AA\ region.

\item  Of the five known ultraviolet coronal forbidden lines, \ion{Fe}{18} $\lambda$974 has important advantages as a spectroscopic diagnostic.  Unlike \ion{Fe}{21} $\lambda$1354 and \ion{Fe}{19} $\lambda$1118, \ion{Fe}{18} is not blended with low temperature lines that can corrupt the intrinsic line profile.  Likewise, the stellar continuum is almost nonexistent at the location of \ion{Fe}{18}, whereas, at longer wavelengths it can create difficulties for isolating the weak coronal forbidden emission.  Although {\em FUSE} does not have an internal wavelength calibration system, LISM absorption in the strong \ion{C}{3} 977.02~\AA\ line, just 660~km~s$^{-1}$ redward of \ion{Fe}{18}, often can be used as a wavelength fiducial (in those cases where $v_{\rm LISM}$ is known independently).  

\item We find that a single Gaussian component is successful for modeling the \ion{Fe}{18} $\lambda$974 profile for all single stars.  The line is fully resolved, so instrumental broadening is negligible.  

\item  The centroid of \ion{Fe}{18} $\lambda$974 is approximately at rest relative to the radial velocity of the star, in contrast to lower temperature ($T~\sim~10^5$~K) transition region lines which are often redshifted 10-15~km~s~$^{-1}$ \citep{wood97,peter99,redfield03}.  Therefore, the high temperature plasma likely is confined in stable magnetic structures; not participating in any large systematic mass motions.  

\item  For all of the slowly rotating stars, thermal broadening dominates the widths of the coronal forbidden lines, consistent with predicted formation temperatures $\sim~6~\times~10^6$~K.  Turbulent broadening appears to be negligible, which is additional evidence that the hot coronal material might be stably entrained in coronal magnetic structures in slowly rotating stars.  

\item  Three of the fastest rotating stars in our sample (AB~Dor, 31~Com, and the G1 star of Capella), all show excess broadening in \ion{Fe}{18} $\lambda$974.  Owing to the overwhelming similarity of \ion{Fe}{18} line widths in all of the slowly rotating stars of our sample, we attribute the excess to ``super-rotational'' broadening.  Namely, the hot gas is co-rotating in extended coronal structures rising well above the stellar surface.  We estimate that this material is located approximately $\Delta$R~$\sim~0.4$-$1.3~$R$_{\star}$ beyond the stellar photosphere.  High speed flows could also potentially explain the excess broadening, although it is not clear why such flows would be observed in the fastest rotating stars.  A detailed discussion concerning the cause of the excess broadening is given in Section~\ref{secfe18}.

\item  The \ion{Fe}{18} $\lambda$974 profile of Capella is complicated by contributions from both stars of the binary.  As demonstrated with the \ion{Fe}{21} $\lambda$1354 line, the relative contribution of each star can vary \citep{linsky98, johnson02, ayres03}.  Using three different observations, at quadrature and conjunction, we find that the G8 primary contributes $\sim~70$\% of the \ion{Fe}{18} emission, and the G1 secondary $\sim~30$\%.  

\item  We find an excellent correlation between \ion{Fe}{18} $\lambda$974 and the 0.2-2.0~keV {\em ROSAT} soft X-ray flux.  Although upper limits mask any trends at lower activity levels, we expect that the correlation will exhibit a ``break'' at a critical activity level where the \ion{Fe}{18} flux will drop faster than the unit slope power law for stars of progressively lower activity.  In particular, \citet{gudel97} have demonstrated that the coronal temperature drops with decreasing activity, and eventually the coronal emission measures will peak below the \ion{Fe}{18} formation temperature, suppressing the forbidden line flux relative to the soft X-ray band.  

\end{itemize}

\acknowledgments
This work is based on data obtained for the Guaranteed Time Team by the
NASA-CNES-CSA {\em FUSE} mission, which is operated by The Johns Hopkins
University. Financial support to U. S. participants has been provided by NASA
contract NAS5-32985, as well as individual {\em FUSE} Guest Investigator grants. This research has made use of the SIMBAD database,
operated at CDS, Strasbourg, France, and the {\em ROSAT} X-ray archive at the HEASARC of the NASA Goddard Space Flight Center.

\clearpage

\clearpage

\clearpage

\clearpage

\clearpage

\clearpage

\clearpage
 
\doublespace
\clearpage
\begin{center}
\begin{deluxetable}{lccccccccc}
\setlength{\tabcolsep}{0.06in} 
\tabletypesize{\scriptsize}
\tablewidth{0pt}
\tablecaption{Properties of Stars Observed by {\em FUSE}\label{properties}}
\tablehead{Star & HD & Spectral & $v_{\rm rad}$ & $m_{\rm V}$ & $B-V$ & $f_{\rm bol}$ & Distance\tablenotemark{a} & $v\sin i$\tablenotemark{b} & Reference \\
Name & & Type & (\kms)& & & ($10^{-7}$ ergs cm$^{-2}$ s$^{-1}$) & (pc) & (\kms) & }
\startdata
              & 23628  & A4 V  & 0 & 7.71 & 0.18 & 0.20 & 130 & 215 & 1 \\
HR~5999       & 144668 & A7 IV & --1.9 & 6.98 & 0.29 & 0.39 & 210 & 204 & 2 \\
$\alpha$ Aql  & 187642 & A7V   &--26.1& 0.77& 0.22& 118 & 5.15 & 210 & 3 \\
$\alpha$ Car  & 45348  & F0 II & 20.5 & --0.72 & 0.15 & 478 & 95.8 & 9 & 2 \\
47 Cas        & 12230  & F0 V  & --26 & 5.28 & 0.30 & 1.89 & 33.6 & 62.1 & 4 \\
$\alpha$ CMi  & 61421  & F5IV-V&--3.2& 0.34& 0.40& 185 & 3.50 & 5.0 & 5 \\
$\upsilon$ Peg& 220657 & F8 IV & --11.1 & 4.40 & 0.61 & 5.2 & 53 & 33.7 & 6 \\
EK Dra        & 129333 & F8    & --30.5 & 7.61 & 0.59 & 0.24 & 33.9 & 17.3 & 7 \\
31 Com        & 111812 & G0 IIIp& --1.4 & 4.94 & 0.63 & 3.4 & 94.2 & 66.5 & 6 \\
HR 9024       & 223460 & G1 III& 0.7 & 5.90 & 0.79 & 1.6 & 130 & 21.5 & 6 \\
$\alpha$ Aur Ab& 34029 & G1III & 52.4, 51.3, 25.4\tablenotemark{c}& 0.76 & 0.65 & 138 & 12.9 & 36 & 8 \\
$\beta$ Dra   & 159181 & G2 Iab& --20.0 & 2.79 & 0.98 & 24 & 110.9 & 10.7 & 6 \\
$\alpha$ Aqr  & 209750 & G2 Ib & 7.5 & 2.95 & 0.96 & 20 & 230 & 7.8 & 6 \\
\aca          & 128620 & G2V   &--24.6 & -0.01 & 0.71 & 283 & 1.35 & 3 & 9 \\
DK UMa        & 82210  & G4 III& --27.2 & 4.57 & 0.75 & 4.6 & 32.4 & 5.5 & 6 \\
$\mu$ Vel     & 93497  & G5 III& 6.2 & 2.72 & 0.91 & 28 & 35.5 & 6.4 & 6 \\
$\alpha$ Aur Aa& 34029 & G8III & 7.1, 8.1, 32.8\tablenotemark{c}& 0.91 & 0.88 & 140 & 12.9 & 3 & 8 \\
$\beta$ Cet   & 4128   & K0III & 13.0 & 2.04 & 1.02 & 56 & 29.4 & 3.0 & 10 \\ 
$\beta$ Gem   & 62509  & K0 IIIb& 3.3 & 1.15 & 1.00 & 121 & 10.3 & $<$ 1.0 & 6 \\
AB Dor        & 36705 & K1V   &29.2& 6.93& 0.80 & 0.51 & 14.9 & 91 & 11 \\
\acb          & 128621 & K2V   &--20.7& 1.33 & 0.88 & 93 & 1.35 & 2 & 12 \\
\eeri         & 22049 & K2V   & 15.5 & 3.73 & 0.88 & 10.4 & 3.22 & 2.0 & 13 \\
$\alpha$ Tau  & 29139  & K5 III& 54.3 & 0.85 & 1.54 & 343 & 20.0 & 2.0 & 6 \\
TW Hya        &        & K8 V  & 13.5 & 11.1 & 0.7 & 0.010 & 56 & $<$ 15 & 14 \\
AU Mic        & 197481 & M0V   & --5.7\tablenotemark{d} & 8.61& 1.44& 0.3 & 10.0 & $<$ 6.2 & 15 \\
$\alpha$ Ori  & 39801  & M2 Iab& 21.0 & 0.58 & 1.77 & 1065 & 131.1 & $<$ 11 & 16 \\
Proxima Cen   &        & M5.5 V& --15.7 & 11.05 & 1.97 & 0.3 & 1.295 & 1.4 & 17 \\
\enddata
\tablenotetext{a}{Hipparcos parallaxes taken from the SIMBAD database.}
\tablenotetext{b}{Source of $v\sin i$ given in Reference column.}
\tablenotetext{c}{Radial velocities calculated for orbital phases (0.34, 0.35, and 0.98, respectively) at the time of the three {\em FUSE} observations \citep{hummel94}.}
\tablenotetext{d}{Based on 37 C~I line observed with STIS \citep{pagano00}.}
\tablerefs{(1) Simon, Drake, \& Kim 1995; (2) Royer et al. 2002; (3) van Belle et al. 2001; (4) Audard et al. 2000; (5) Hale 1994; (6) De Medeiros \& Mayor 1999; (7) Montes et al. 2001; (8) Strassmeier et al. 1988; (9) Pallavicini et al. 1981; (10) Gray 1989; (11) Brandt et al. 2001; (12) Smith, Edvardsson, \& Frisk 1986; (13) Fekel 1997; (14) Franchini et al. 1992; (15) Pettersen 1980; (16) Gray 2000; (17) Wood et al. 1994.}
\end{deluxetable}
\end{center}


\clearpage
\begin{center}
\begin{deluxetable}{lccccc}
\tabletypesize{\tiny}
\tablewidth{0pt}
\tablecaption{Summary of {\em FUSE} Observations\label{summary}}
\tablehead{Star  &  Data  &  Obs.  &  Aperture & Number & Exposure \\
Name  &  Set  &  Date  &    & Exposures & Time (ks) }
\startdata
HD 23628     & P1860501  & 2000 Dec 29 & LWRS &\phn 6 &\phn 6.9 \\
HD 144668    & P1860201  & 2001 Aug 20 & LWRS &\phn 4 &\phn 4.8 \\
$\alpha$ Aql & P1180701  & 2001 Sep 14 & LWRS &\phn 8 &\phn 4.2      \\
$\alpha$ Car & P1180101  & 2000 Dec 11 & LWRS &\phn 2 &\phn 5.6 \\
$\alpha$ Car & P2180101  & 2001 Oct 25 & LWRS &\phn 3 &\phn 6.1 \\
$\alpha$ Car & P2180102  & 2001 Oct 26 & LWRS &\phn 9 & 11.1 \\
47 Cas       & P1860601  & 2000 Nov 26 & LWRS &\phn 3 & 10.9 \\
Procyon      & P1041801  & 2001 Oct 21 & MDRS &\phn 7 &\phn 5.3      \\
$\upsilon$ Peg&B0720301  & 2001 Jul  1 & LWRS &\phn 4 & 11.8 \\
EK Dra       & P1860401  & 2000 Mar  1 & LWRS &\phn 3 & 10.3 \\
31 Com       & P1180401  & 2001 Apr 20 & LWRS &\phn 8 & 12.9 \\
HR 9024      & B0720101  & 2001 Jul 10 & LWRS & 19    & 37.2 \\
$\alpha$ Aur & P1040301  & 2000 Nov  5 & LWRS & 10    & 14.2 \\
$\alpha$ Aur & P1040302  & 2000 Nov  7 & LWRS & 10    & 12.3 \\
$\alpha$ Aur & P1040303  & 2001 Jan 11 & MDRS & 28    & 21.4 \\
$\beta$ Dra  & P1180301  & 2000 May  9 & LWRS &\phn 2 &\phn 5.6 \\
$\beta$ Dra  & P2180301  & 2001 Jun 30 & LWRS &\phn 8 & 16.4 \\
$\alpha$ Aqr & P1180201  & 2000 Jun 29 & LWRS &\phn 1 &\phn 3.3 \\
$\alpha$ Aqr & P1180202  & 2001 Oct  8 & LWRS &\phn 7 & 15.9 \\
$\alpha$ Aqr & P2180201  & 2001 Jun 16 & LWRS & 11    & 34.4 \\
$\alpha$ Aqr & P2180202  & 2001 Oct  7 & LWRS &\phn 5 & 10.5 \\
\aca         & P1042601  & 2001 Jun 25 & MDRS &\phn 8 & 15.3 \\
DK UMa       & C1070102  & 2002 Feb 14 & LWRS &\phn 6 & 24.6 \\
$\mu$ Vel    & C1070201  & 2002 Mar  5 & LWRS &\phn 8 & 25.2 \\
$\beta$ Cet  & P1180501  & 2000 Dec 10 & LWRS & 10    & 12.6 \\
$\beta$ Gem  & P1180601  & 2000 Nov 11 & LWRS &\phn 7 & 21.9 \\
AB Dor       & X0250201\tablenotemark{a}&1999 Oct 20 & LWRS &\phn 1 & 22.1 \\
AB Dor       & X0250203  & 1999 Dec 14 & LWRS &\phn 6 & 24.2 \\
\acb         & P1042501  & 2001 Jun 24 & MDRS & 13    & 22.7 \\
\eeri        & P1040701  & 2000 Dec  8 & MDRS & 15    & 34.8 \\
$\alpha$ Tau & P1040901  & 2001 Jan 14 & MDRS &\phn 8 & 12.2 \\
TW Hya       & P1860101  & 2000 Jun  3 & LWRS &\phn 3 &\phn 2.1 \\
AU Mic       & P1180801  & 2000 Aug 26 & LWRS &\phn 9 & 17.3        \\
AU Mic       & P2180401  & 2001 Oct 10 & LWRS & 13    & 26.5        \\
$\alpha$ Ori & P1180901  & 2000 Nov  3 & LWRS &\phn 7 & 10.4 \\
Proxima Cen  & P1860701  & 2000 May 16 & LWRS &\phn 2 & 6.9 \\
\enddata
\tablenotetext{a}{The star was only observed in the LiF channels (987-1187~\AA).}
\end{deluxetable}
\end{center}

\clearpage
\begin{center}
\begin{deluxetable}{lcccc}
\tabletypesize{\scriptsize}
\tablewidth{0pt}
\tablecaption{Observed Iron and X-ray Fluxes\label{observed}}
\tablehead{Star & \multicolumn{3}{c}{Integrated Flux ($10^{-14}$ ergs cm$^{-2}$ s$^{-1}$)} & Reference \\
Name & Fe~XVIII\tablenotemark{a} & Fe~XIX\tablenotemark{a,b} & X-ray & \\
 & $\lambda$974 & $\lambda$1118 & (0.2-2.0 keV) & }
\startdata
HD 23628       & $<$ 1.8 & $<$ 1.0 & 90 $\pm$ 20\tablenotemark{c} & 1 \\
HD 144668      & $<$ 1.2 & $<$ 3.0 & 76 $\pm$ 4\tablenotemark{c} & 1 \\
$\alpha$ Aql   & $<$ 3.1 & $<$ 7.7 & 71 $\pm$ 2 & 1 \\
$\alpha$ Car   & $<$ 1.5 & $<$ 1.3 & 420 $\pm$ 10 & 1 \\
47 Cas         & 1.0 $\pm$ 0.3 & 1.2 $\pm$ 0.2 & 1310 $\pm$ 30 & 1 \\
$\alpha$ CMi   & $<$ 3.0 & $<$ 4.9 & 1300 $\pm$ 20 & 1 \\
$\upsilon$ Peg & $<$ 1.0 & $<$ 0.6 & 600 $\pm$ 100 & 2 \\
EK Dra         & 1.0 $\pm$ 0.3 & 0.6 $\pm$ 0.1 & 730 $\pm$ 190 & 1 \\
31 Com         & 0.8 $\pm$ 0.3 & 1.5 $\pm$ 0.2 & 800 $\pm$ 100 & 2 \\
HR 9024        & $<$ 0.8 & $<$ 0.6 & 1000 $\pm$ 500 & 2 \\
$\beta$ Dra    & $<$ 0.9 & $<$ 1.1 & 190 $\pm$ 10 & 1 \\
$\alpha$ Aqr   & $<$ 0.5 & $<$ 0.3 & $<$ 20 & 1 \\
\aca           & $<$ 1.1 & $<$ 2.4 & 700 $\pm$ 200 & 2 \\
DK UMa         & 1.1 $\pm$ 0.2 & 1.0 $\pm$ 0.1 & 1400 $\pm$ 100 & 2 \\
$\mu$ Vel      & 1.4 $\pm$ 0.2 & 1.9 $\pm$ 0.2 & 1600 $\pm$ 100 & 2 \\
$\alpha$ Aur   & 11.5 $\pm$ 1.3 & 31.0 $\pm$ 4.7 & 12700 $\pm$ 1600 & 2 \\
$\beta$ Cet    & 3.8 $\pm$ 0.4 & 4.1 $\pm$ 0.4 & 2000 $\pm$ 200 & 2 \\
$\beta$ Gem    & $<$ 0.9 & $<$ 0.6 & 30 $\pm$ 2 & 1 \\
AB Dor         & 1.6 $\pm$ 0.3 & 1.6 $\pm$ 0.2 & 5500 $\pm$ 2000 & 1 \\
\acb           & $<$ 1.3 & $<$ 2.4 & $\sim$ 1200 & 1 \\
\eeri          & 0.8 $\pm$ 0.2 & 0.7 $\pm$ 0.1 & 1500 $\pm$ 500 & 2 \\
$\alpha$ Tau   & $<$ 0.9 & $<$ 0.6 & $<$ 4 & 2 \\
TW Hya         & $<$ 1.8 & $<$ 0.5 & 300 $\pm$ 10 & 1 \\
AU Mic         & 0.6 $\pm$ 0.1 & 0.7 $\pm$ 0.1 & 2200 $\pm$ 1000 & 1 \\
$\alpha$ Ori   & $<$ 0.8 & $<$ 0.6 & $<$ 13 & 1 \\
Proxima Cen    & $<$ 0.4 & $<$ 0.5 & 700 $\pm$ 200 & 2 \\
\enddata
\tablenotetext{a}{Fluxes calculated by direct numerical integration of fixed $\pm$~150~km~s$^{-1}$ region for all stars with $v\sin~i~<~50$~km~s$^{-1}$, and $\pm$~250~km~s$^{-1}$ for all stars with $v\sin~i~>~50$~km~s$^{-1}$.}
\tablenotetext{b}{Fluxes include contributions from possible blends with low temperature C~I lines.}
\tablenotetext{c}{Hard X-ray source: probably not the A star, but an unseen companion.}
\tablerefs{(1) X-ray fluxes calculated from 0.2-2.0~keV band from {\em ROSAT} WGACAT (from HEASARC); (2) Ayres et al. 2003.}
\end{deluxetable}
\end{center}

\clearpage
\begin{center}
\begin{deluxetable}{lcccccc}
\tabletypesize{\footnotesize}
\tablewidth{0pt}
\tablecaption{Fe~XVIII Emission Line Fit Parameters\label{fits}}
\tablehead{Star&HD&$v$\tablenotemark{a} & Integrated Flux & FWHM$_{\rm obs}$\tablenotemark{b} & FWHM\tablenotemark{c} & $\log T_{\rm eq}$\tablenotemark{d} \\
Name & & (\kms) & (10$^{-14}$ ergs cm$^{-2}$ s$^{-1}$) & (\kms) & (\kms) & (K) }
\startdata
47 Cas & 12230 & --4 $\pm$ 15 & 1.0 $\pm$ 0.1 & 113 $\pm$ 10 & 87 $\pm$ 10 & $6.96^{+0.07}_{-0.08}$ \\
EK Dra & 129333 & --3 $\pm$ 15 & 0.96 $\pm$ 0.10 & 78 $\pm$ 10 & 74 $\pm$ 10 & $6.82^{+0.08}_{-0.09}$ \\
31 Com & 111812 & 22 $\pm$ 35 & 0.70 $\pm$ 0.07 & 149 $\pm$ 14 & 125 $\pm$ 14 & $7.28^{+0.06}_{-0.07}$\tablenotemark{e} \\
DK UMa & 82210 & --4 $\pm$ 10 & 1.19 $\pm$ 0.13 & 75.3 $\pm$ 9.6 & 75.3 $\pm$ 9.6 & $6.84^{+0.07}_{-0.09}$ \\
$\mu$ Vel & 93497 & --2.6 $\pm$ 5.0 & 1.5 $\pm$ 0.2 & 77 $\pm$ 10 & 77 $\pm$ 10 & $6.86^{+0.07}_{-0.09}$ \\
$\beta$ Cet & 4128 & 1.1 $\pm$ 5.0 & 3.53 $\pm$ .35 & 83.4 $\pm$ 6.8 & 83.4 $\pm$ 6.8 & 6.93 $\pm$ 0.05 \\
AB Dor & ~36705& --16 $\pm$ 19 & 1.92 $\pm$ 0.32 & 309 $\pm$ 35 & 288 $\pm$ 35 & $8.00^{+0.07}_{-0.08}$\tablenotemark{f} \\
\eeri  & ~22049& 7 $\pm$ 10 & 0.97 $\pm$ 0.10 & 75.8 $\pm$ 8.8 & 75.8 $\pm$ 8.8 & $6.84^{+0.07}_{-0.08}$ \\
AU Mic & 197481& --5 $\pm$ 10 & 0.56 $\pm$ 0.08 & 81 $\pm$ 14 & 81 $\pm$ 14 & $6.90^{+0.09}_{-0.12}$ \\
\enddata
\tablenotetext{a}{Relative to the radial velocity of the stellar photosphere.}
\tablenotetext{b}{Observed FWHM of the Fe~XVIII feature.}
\tablenotetext{c}{FWHM after deconvolving the photospheric rotational velocity; negligible for $v\sin~i~\leq~10$~km~s$^{-1}$.}
\tablenotetext{d}{Equivalent Temperature required to account for deconvolved FWHM, assuming no turbulent velocity contribution; expected temperature range for the formation of Fe~XVIII is $6.62~\leq~\log~T~({\rm K})~\leq~6.93$.}
\tablenotetext{e}{$T_{\rm eq}$ too large; can be explained with rotational velocity $v\sin~i~\sim~95^{+80}_{-20}$~km~s$^{-1}$, or extended emission at  $\Delta$R~$\sim~0.4^{+1.2}_{-0.3}$~R$_{\star}$ above the stellar photosphere.}
\tablenotetext{f}{$T_{\rm eq}$ too large; can be explained with rotational velocity $v\sin~i~\sim~210^{+60}_{-40}$~km~s$^{-1}$, or extended emission at  $\Delta$R~$\sim~1.3^{+0.7}_{-0.4}$~R$_{\star}$ above the stellar photosphere.}
\end{deluxetable}
\end{center}

\clearpage
\begin{center}
\begin{deluxetable}{lccccccccc}
\rotate
\tabletypesize{\footnotesize}
\tablewidth{0pt}
\tablecaption{Capella Fe~XVIII Emission Line Fit Parameters\label{fitscapella}}
\tablehead{Star&HD&Date&Phase&$v$\tablenotemark{a} & Integrated Flux & FWHM$_{\rm obs}$\tablenotemark{b} & FWHM\tablenotemark{c} & $\log T_{\rm eq}$\tablenotemark{d} & $\chi^2_{\nu}$ \\
Name & & & & (\kms) & (10$^{-14}$ ergs cm$^{-2}$ s${-1}$) & (\kms) & (\kms) & (K) & }
\startdata
\\[-7.75ex]
\multicolumn{9}{c}{Single Gaussian Fit} \\
\hline
$\alpha$ Aur (G8) & 34029 & 2000 Nov 5 & 0.34 & 5.5 $\pm$ 5.0 & 12.7 $\pm$ 1.3 & 97.5 $\pm$ 4.3 & 97.5 $\pm$ 4.3 & 7.06 $\pm$ 0.03 & 3.31 \\
$\alpha$ Aur (G8) & 34029 & 2000 Nov 7 & 0.35 & 3.0 $\pm$ 5.0 & 9.76 $\pm$ 0.98 & 92.8 $\pm$ 6.8 & 92.8 $\pm$ 6.8 & $7.02^{+0.04}_{-0.05}$ & 6.21 \\
$\alpha$ Aur (G8) & 34029 & 2001 Jan 11 & 0.98 & --8.1 $\pm$ 5.0 & 9.62 $\pm$ 0.96 & 80.1 $\pm$ 5.0 & 80.1 $\pm$ 5.0 & 6.89 $\pm$ 0.04 & 9.99 \\ 
\hline
\multicolumn{9}{c}{Double Gaussian Fit\tablenotemark{e}} \\
\hline
$\alpha$ Aur (G8) & 34029 & 2000 Nov 5 & 0.34 & 0.0 & 9.09 $\pm$ 0.91 & 80.1 & 80.1 & 6.89 & 1.12 \\
\phantom{$\alpha$ Aur }(G1) & & & & 0.0 & 4.44 $\pm$ 0.72 & 169 $\pm$ 25 & 164 $\pm$ 25 & 7.51$^{+0.08}_{-0.11}$\tablenotemark{f} & 1.12 \\
$\alpha$ Aur (G8) & 34029 & 2000 Nov 7 & 0.35 & 0.0 & 7.47 $\pm$ 0.75 & 80.1 & 80.1 & 6.89 & 1.14 \\
\phantom{$\alpha$ Aur }(G1) & & & & 0.0 & 2.86 $\pm$ 0.66 & 158 $\pm$ 28 & 151 $\pm$ 30 & 7.44$^{+0.11}_{-0.14}$\tablenotemark{f} & 1.14 \\
$\alpha$ Aur (G8) & 34029 & 2001 Jan 11 & 0.98 & 0.0 & 7.12 $\pm$ 0.62 & 80.1 & 80.1 & 6.89 & 2.27 \\
\phantom{$\alpha$ Aur }(G1) & & & & 0.0 & 2.15 $\pm$ 0.34 & 164 & 158 & 7.48\tablenotemark{f} & 2.27 \\
\enddata
\tablenotetext{a}{Relative to the radial velocity of the stellar photosphere.}
\tablenotetext{b}{Observed FWHM of the Fe~XVIII feature.}
\tablenotetext{c}{FWHM after deconvolving the photospheric rotational velocity; negligible for $v\sin~i~\leq~10$~km~s$^{-1}$.}
\tablenotetext{d}{Equivalent Temperature required to account for FWHM, assuming no turbulent velocity contribution; expected temperature range for the formation of Fe~XVIII is $6.62~\leq~\log~T~({\rm K})~\leq~6.93$.}
\tablenotetext{e}{Only values with error bars are free parameters in double Gaussian fits.}
\tablenotetext{f}{$T_{\rm eq}$ too large; can be explained with rotational velocity $v\sin~i~\sim~80$~km~s$^{-1}$, or extended emission at $\Delta$R~$\sim$~1.2~R$_{\star}$ above the stellar photosphere.}
\end{deluxetable}
\end{center}

\end{document}